# Performance evaluation of job schedulers on Hadoop YARN

Jia-Chun Lin
Department of Informatics, University of Oslo
Gaustadallèen 23 B, Oslo, N-0373, Norway
kellylin@ifi.uio.no

Ming-Chang Lee
Department of Communication Systems, Simula Research Laboratory
Martin Linges vei 25, Fornebu, 1364, Norway
mclee@simula.no

September 10, 2018

# Performance Evaluation of Job Schedulers on Hadoop YARN

Jia-Chun Lin[1] and Ming-Chang Lee[2]

[1]Department of Informatics, University of Oslo, Norway
[2]Department of Communication Systems, Simula Research Laboratory, Norway
{kellylin1219, mingchang1109}@gmail.com

**Abstract**

To solve the limitation of Hadoop on scalability, resource sharing, and application support, the open source community proposes the next generation of Hadoop's compute platform called YARN (Yet Another Resource Negotiator) by separating resource management functions from the programming model. This separation enables various application types to run on YARN in parallel. To achieve fair resource sharing and high resource utilization, YARN provides the capacity scheduler and the fair scheduler. However, the performance impacts of the two schedulers are not clear when mixed applications run on a YARN cluster. Therefore, in this paper, we study four scheduling-policy combinations (SPCs for short) derived from the two schedulers and then evaluate the four SPCs in extensive scenarios, which consider not only four application types, but also three different queue structures for organizing applications. The experimental results enable YARN managers to comprehend the influences of different SPCs and different queue structures on mixed applications. The results also help them to select a proper SPC and an appropriate queue structure to achieve better application execution performance.

**Keywords**: Hadoop, YARN, capacity scheduler, fair scheduler, queue structure, performance evaluation.

## 1. Introduction

Hadoop [1] is an open-source software framework supported by Apache to process high volume of datasets on a cluster comprising large number of commodity machines. Because of its simplicity, cost efficiency, scalability, and fault tolerance, a wide variety of organizations and companies, such as Google, Yahoo!, Facebook, and Amazon, have used Hadoop for both research and production [2]. However, the original Hadoop has several limitations [3]. One example is that the slot-based resource allocation for map tasks and reduce tasks bottlenecks the resource of an entire Hadoop cluster and results in low resource utilization [3]. Another example is that the original Hadoop supports only one type of programming model, i.e., MapReduce [4], which is not suitable for processing all kinds of large-scale computations [3][5][6].

To solve these limitations, the open-source community introduced the next generation of Hadoop's compute platform called YARN (which is short for Yet Another Resource Negotiator) [3]. Another names are MapReduce 2.0 and MRv2. YARN allows individual applications to utilize the resources of a cluster in a shared and multi-tenant manner. Different from the original Hadoop (i.e., all versions before MRv2), YARN separates resource management functions from the programming model, and therefore can support not only MapReduce but also other programming models, including Spark [5], Storm [7], Tez [8], and REEF [9]. In other words, this separation enables that various types of applications can execute on YARN in parallel.

To enable a shared compute environment, YARN provides two schedulers to schedule resources to applications. One is the capacity scheduler (the default scheduler on YARN) [10], and the other is the fair scheduler [11]. Both of them can organize application submissions into a queue hierarchy. However, the former guarantees a minimum amount of resources for each queue and uses FIFO to schedule applications within a leaf queue. The latter fairly shares resources among all queues and offers three policies, including FIFO, Fair, and Dominant Resource Fairness (DRF for short) [12], to share resources for all running applications within a queue. All of the abovementioned scheduling approaches form the following four scheduling-policy combinations (SPCs for short) and provide great flexibility for YARN managers to achieve their goals, such as fair resource sharing and high resource utilization.

1. Cap-FIFO, i.e., the capacity scheduler with the FIFO scheduling policy.
2. Fair-FIFO, i.e., the fair scheduler with the FIFO scheduling policy.
3. Fair-Fair, i.e., the fair scheduler with the fair scheduling policy.
4. Fair-DRF, i.e., the fair scheduler with the DRF scheduling policy.

Although YARN supports the four SPCs and diverse application types, it is unclear how these SPCs perform when they are individually used to schedule mixed applications. Besides, their performances are also unknown when different queue structures are utilized. Hence, in this paper, we survey the four SPCs and all programming models supported by YARN, and then classify all applications into several types. After that, we conduct extensive experiments to evaluate and compare the performance impacts of the four SPCs on diverse metrics by considering not only a workload consisting of mixed application types, but also the following three scenarios. The purpose is to study whether queue structures influence the performances of the four SPCs or not.

1. One-queue scenario: In this scenario, there is only one queue in our YARN cluster. Hence, all application submissions must wait in this queue before they are executed.
2. Separated-queue scenario: In this scenario, each type of applications is individually put into a separate queue.
3. Merged-queue scenario: In this scenario, there are two queues. One is for applications that will eventually stop by themselves. The other queue is for the rest of applications.

The experimental results show that (1) All SPCs suffer from a resource fragmentation problem, which will be explained later. This problem causes that none of the SPCs could successfully complete a workload consisting of mixed applications; (2) None of the four SPCs always has the best application execution performance in all scenarios; (3) Among the three scenarios, employing the merged-queue scenario is the most appropriate for all SPCs since they can achieve a higher workload completion rate and a shorter workload turnaround time than they are in the other two scenarios.

The contributions of this paper are as follows. (1) This paper provides a comprehensive survey on current schedulers, SPCs, programming models, and application types supported by YARN; (2) We extensively evaluate and compare the four SPCs by considering not only mixed application types, but also diverse queue-structure scenarios; (3) Based on our experimental results, YARN managers can choose an appropriate SPC and queue structure to achieve a better application performance for their YARN clusters.

The rest of this paper is organized as follows. Section 2 describes the related work. Section 3 surveys the origin of YARN. Section 4 introduces the two schedulers supported by YARN and the four SPCs derived from the two schedulers. Section 5 describes the programming models supported by YARN and applications that each programming model can best express and process. In Section 6, extensive experiments are conducted and experimental results are discussed. Section 7 concludes this paper and outlines our future work.

## 2. Related Work

There have been several survey articles on job scheduling in Hadoop. Rao and Reddy [13] studied various Hadoop schedulers, including the default FIFO scheduler [4], the fair scheduler, the capacity scheduler, and the delay scheduling [14] etc., by summarizing the advantages and disadvantages of these schedulers. However, the authors only introduced those schedulers without conducting any experiments to evaluate and compare their performances. Kulkarni and Khandewal [15] also surveyed several job scheduling algorithms in Hadoop. But similar to [13], no performance evaluation was presented in [15]. Another related survey paper can be found in [16].

In order to improve Hadoop scheduling in terms of job completion time, data locality, or other performance metrics, many researchers have introduced their scheduling algorithms for Hadoop and performed experiments to compare their algorithms with those used by Hadoop, e.g., FIFO, the capacity scheduler, and the fair scheduler. For example, Zaharia et al. [14] proposed the delay scheduling algorithm to improve data locality while maintaining fairness. The authors evaluated their algorithm with the default FIFO scheduler and the fair scheduler, and demonstrated that their algorithm outperforms the others in terms of data locality and job response time. The context aware scheduler proposed by Kumar et al. [17] and the ThroughputScheduler presented by Gupta et al. [18] are two examples for improving performance on heterogeneous Hadoop clusters. Both of them were designed to assign tasks to the most capable nodes such that the resource requirements of the tasks can be satisfied. They also evaluated their schedulers with those used by Hadoop. However, the evaluation in [17] is based on a simulation, rather than a real experiment done in [18]. Lee et al. [19] improved data locality for both map and reduce tasks, avoid job starvation, and improve job execution performance by introducing JoSS (which stands for hybrid job-driven scheduling scheme). Two variations of JoSS were further introduced to separately achieve a better map-data locality and a faster task assignment. The authors conduct extensive experiments to evaluate and compare the two variations with current scheduling algorithms supported by Hadoop. Different from all above studies, in this paper, we focus on studying the performance impacts of different scheduling-policy combinations supported by YARN on mixed applications.

Other studies have been presented to study the performance of Hadoop from different perspectives. Gu and Li in [20] evaluated the performances of Hadoop and Spark in terms of time and memory cost when running iterative operations. Their results show that Spark performs faster than Hadoop, but it consumes more memory than Hadoop. Hence, if memory is insufficient at a moment, the speed advantage of Spark will reduce. Xavier et al. [21] presented performance comparison between the current container-based systems, including Linux VServer, OpenVZ, and Linux Containers (LXC), for MapReduce clusters. Lin et al. [22] studied the impact of various MapReduce policies on job completion reliability and job energy consumption. To our best knowledge, the study presented in this paper is the first one that comprehensively studies the impact of current scheduling-policy combinations supported by YARN on various applications types and meanwhile takes different queue structures into account.

## 3. The Origin of YARN

In this section, we briefly describe the original Hadoop and its limitations, and then introduce how YARN solves these limitations.

### 3.1 Hadoop

Hadoop [1] mainly consists of two key components: Hadoop distributed file system (HDFS) and MapReduce. The former is designed to reliably store large files across machines in a large cluster by splitting each file into several blocks and replicating each block to several machines. The latter is a distributed programming model for users to simply specify their jobs as two primitive functions (i.e., Map and Reduce) without requiring to handle resource management, job scheduling, and fault tolerance [4]. Fig. 1 illustrates the execution flow of a MapReduce job on Hadoop. First, a client submits a job to JobTracker, which is a master server responsible to coordinate and schedule the execution of all jobs. Next, JobTracker schedules each task of the job to an available slave node called TaskTracker. Each TaskTracker provides a fixed number of map slots and reduce slots to respectively execute map tasks and reduce tasks assigned by JobTracker. During the execution of the job, JobTracker not only monitors task progress but also provides fault tolerance for each failed task. When all tasks of the job are completed, JobTracker informs the client about the completion.

The design of Hadoop leads to several limitations on availability, scalability, resource utilization, and application support [3]. First, JobTracker is a single point of failure. If it crashes, all jobs cannot proceed and must restart. Second, Hadoop only supports one single type of programming model, i.e., MapReduce. Although MapReduce can express and process many applications, it is unsuitable for iterative applications, streaming applications, interactive data-mining applications, and graph applications [3]. Third, limiting a slot to execute only one type of task (i.e., either a map task or a reduce task) might cause low cluster utilization since map slots might be fully utilized while reduce slots are empty (and vice-versa).

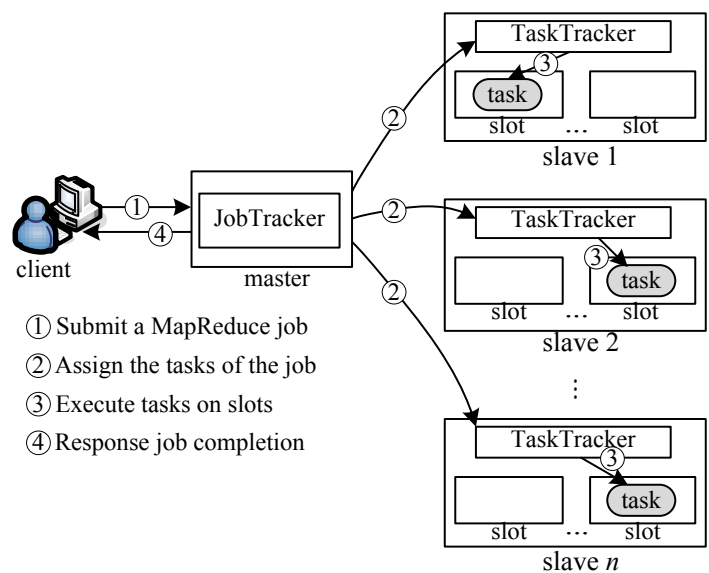

Fig. 1. The execution flow of a MapReduce job on Hadoop.

## 3.2 YARN

To solve the above limitations, YARN separates resource management functions from the programming model and introduces the following components/roles:

- A global Resource Manager (RM for short): It acts as a center authority in a YARN cluster and focuses on scheduling, i.e., tracking resource usage and allocating available resources to applications based on the resource requirements of the applications. Unlike JobTracker, RM does not monitor applications' statuses and restart any failed tasks. This responsibility separation enables RM to improve YARN's scalability.

- A per-slave Node Manager (NM for short): It is an agent in a slave node to report the node's health to RM, wait for instructions from RM, manage containers running on the node, launch containers for applications, and monitor resource usage of individual containers.

- A per-application Application Master (AM for short): It is the head of a job, which requests containers from RM and works with NM to execute and manage the execution flow of the job.

- A per-application container: It is a logical bundle of resources (e.g., 1 GB of

memory, 1 CPU) on a slave node [3]. Unlike map slots and reduce slots used in the original Hadoop, a container can run any type of task. This allows YARN to properly allocate resources to applications and improve resource utilization.

With the above improvements, YARN supports diverse programing models and allows various application types to execute on YARN in parallel. Fig. 2 illustrates the execution flow of an application on YARN. In step 1, a client submits an application to RM. Then RM in step 2 negotiates with a specified container so as to launch the AM of the application on the container. After the AM starts, it registers with RM and starts requesting containers from RM (see step 3). Once receiving a container from RM, the AM in steps 4 and 5 provides the container launch specification to the corresponding NM and executes the application code on the container. During the application execution, the client can directly communicate with the AM to know current progress and status (see step 6). When the application completes, the AM deregisters with RM and releases all containers it uses in step 7. Finally, the AM informs the client about the completion.

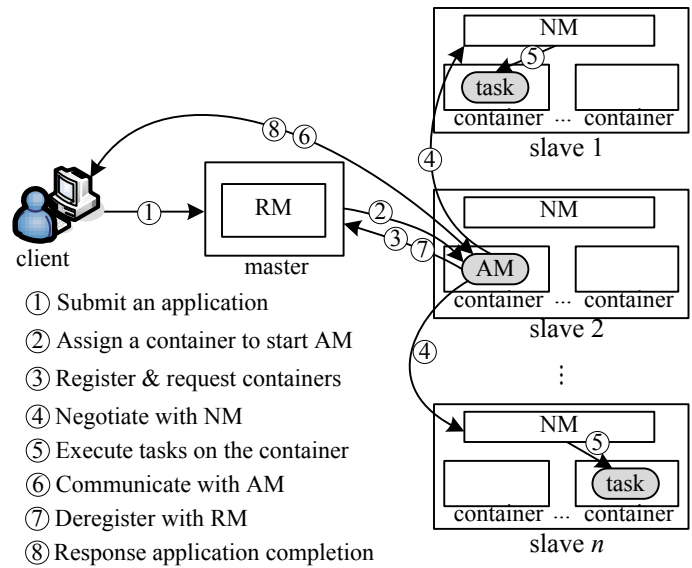

① Submit an application
② Assign a container to start AM
③ Register & request containers
④ Negotiate with NM
⑤ Execute tasks on the container
⑥ Communicate with AM
⑦ Deregister with RM
⑧ Response application completion

Fig. 2. The execution flow of an application on YARN.

## 4. Job Schedulers Supported by YARN

In this section, we describe the main concepts of the capacity scheduler and the fair schedulers, and then introduce four SPCs derived from the two schedulers.

### 4.1 The capacity scheduler

The capacity scheduler [10] is designed for multiple tenants to share a large cluster such that their applications can be allocated resources under constraints of allocated capacities. The capacity scheduler supports hierarchical queues to reflect the structure of organizations/groups that utilize cluster resources. In general, a queue hierarchy contains three types of queues: root, parent, and leaf. Only leaf queues accept application submission. In fact, the root queue represents the cluster itself, rather than any organization/group, whereas a parent queue represents for an organization/group or a sub-organization/sub-group.

The capacity scheduler provides capacity guarantee by allocating a fraction of the cluster resources to each queue. YARN managers can also limit the maximum capacity for each queue. For example, if the minimum and maximum capacity allocation of a queue are respectively 40% and 60%, it means that this queue can use at least 40% and at most 60% of the cluster resources. To provide elasticity, the capacity scheduler allows a queue to use more resource than its capacity allocation if the capacity allocated to the other queues is not fully utilized.

When cluster resource is available, the capacity scheduler works as follows:

Step 1. It calculates the current capacity used by each leaf queue, i.e., the total amount of resources used by all applications in each leaf queue. Then, the scheduler picks up the most under-served queue, i.e., the one with the lowest used capacity among all leaf queues.

Step 2. The scheduler selects an application from the most under-served queue in a FIFO order, i.e., the application that is submitted to the queue first will be allocated resource first.

Step 3. Upon an application is chosen, the resource is further scheduled to a task of the application based on the priorities of resource requests assigned by the application.

In this paper, we use the term "inter-queue scheduling" to represent the process of choosing a leaf queue from all leaf queues, and use the term "intra-queue scheduling" to indicate the process of choosing an application from a leaf queue. Hence, we can see that the capacity scheduler only provides only one SPC, i.e., Cap-FIFO.

## 4.2 The fair scheduler

The fair scheduler [11] aims to assign resources to applications such that these applications get fair resources over time. Similar to the capacity scheduler, the fair scheduler supports hierarchical queues to reflect the structure of an organization/group sharing a cluster, enables each queue to get its guaranteed minimum capacity, and limits the maximum capacity for each queue. However, different from the capacity scheduler, the fair scheduler offers three policies for YARN managers to flexibly share resources to applications within a queue:

1. FIFO: When this is applied to a leaf queue, available resource will be assigned to an application that arrives at this queue first.
2. Fair: When this is applied to a leaf queue, available resource will be allocated to an application that currently uses the least amount of memory among all applications within the queue.
3. Dominant resource fairness (DRF): DRF is a fair resource sharing model introduced by Ghodsi et al. [12] to generalize max-min fairness to multiple resource types. For each user, DRF calculates the share of each resource allocated to the user and considers the maximum one among all the shares as the user's dominant share. The resource corresponding to the dominant share is the user's dominant resource. For example, user $U$ has been allocated <1 CPU, 5000 MB> and the entire capacity of a cluster is <4 CPUs, 8000 MB>, implying that $U$'s current CPU share and memory share are 1/4 and 5/8, respectively. Hence, $U$'s dominant share is 5/8, and $U$'s dominant resource is memory. The goal of DRF is to equalize the dominant shares of all users. When DRF is applied to a leaf queue, available resource will be preferentially allocated to an application that has the smallest dominant share in the queue.

With the above three scheduling policies, the fair scheduler provides three available SPCs, i.e., Fair-FIFO, Fair-Fair, and Fair-DRF. Whenever cluster resource is available, the fair scheduler works as follows:

Step 1. The fair scheduler picks up a leaf queue based on the scheduling policies set for each level of the queue hierarchy. First, it chooses a sub-queue of the root queue, say queue $X$, based on the designate scheduling policy. Next, it chooses a sub-queue of queue $X$ based on the designate scheduling policy. Then it repeats the same procedure until a leaf queue is reached.

Step 2. The scheduler picks up an application from the chosen leaf queue based on the scheduling policy set for the leaf queue.

Step 3. Upon an application is chosen, the resource is further scheduled to a task of the application based on the priorities of resource requests assigned by the application.

## 5. Programming Models and Application Types

In this section, we introduce programming models supported by YARN and classify applications run on YARN based on their properties.

## 5.1 Programming Models

### 5.1.1 MapReduce

MapReduce [4] allows a programmer to express his/her computation as a map function and a reduce function. The former takes an input key/value pair and generates intermediate key/value pairs. The latter merges all intermediate key/value pairs associated with the same key and produces final results. Because of the two functions, the execution of a MapReduce job comprises a map stage and a reduce stage. In the map stage, each map task runs the user-defined map function to process an input-data block and generate intermediate key/value data. In the reduce stage, each reduce task runs the user-defined reduce function to process the intermediate key/value data and produce the final result. It is well known that MapReduce is designed and suitable for batch applications [23], such as log analysis and text processing.

### 5.1.2 Apache Tez

Apache Tez was designed to generalize the MapReduce paradigm. By modeling data processing as a directed acyclic graph (DAG) with vertices representing application logic and edges representing movement of data, Apache Tez allows users to express complex data-processing tasks. When a Tez job executes, it starts at the root vertices of the DAG and continues down the directed edges till reaching the leaf vertices. Only when all the vertices in the DAG are completed, the job is complete.

### 5.1.3 Spark

Spark is an open-source computing framework developed to support applications that cannot be efficiently processed by MapReduce, i.e., the applications that reuse a set of data across multiple parallel operations. Typical examples include iterative machine-learning applications and interactive data-analysis applications. Spark employs an abstraction called a resilient distributed dataset (RDD for short) [5], which is a read-only collection of objects split across multiple machines/nodes. Users can cache a RDD in memory across multiple Spark workers and reuse it by using parallel operations, rather than keep retrieving it from HDFS.

With an advanced DAG engine of Spark, a Spark application can have any number of stages. Furthermore, Spark provides Spark Streaming [24] and GraphX [25]. The former allows users to process live data streams in a high-throughput and fault-tolerant manner, whereas the latter enables users to deal with large-scale graph-parallel computation.

### 5.1.4 Storm

Storm [7] is an open-source distributed computation system for processing large streams of data in real time. In Storm, a stream is an unbounded sequence of tuples. Each tuple is an ordered list of elements, e.g., (3, 2, 5) is a 3-tuple. Each Storm application is defined as a topology to process incoming streams of data. More specifically, a topology is a directed graph with a set of vertices and edges. The vertices could be either spout or blot. A spout reads tuples from an external source and emits them into the topology. A blot processes input streams and generates output streams. A Storm topology does not eventually finish by itself. Instead, it keeps processing incoming streams until it is killed.

### 5.2 Application types

Based on Section 5.1, we learn that YARN supports various applications, and these applications can be further classified into four types:

1. Two-stage: This type refers to all application expressed by MapReduce.
2. DAG: The type refers to all applications that can be expressed as a DAG, regardless of its structure and the number of its stages, vertices, and edges.
3. Directed cycle graph (DCG): This type refers to all graph-parallel computations without a directed cycle.
4. Streaming: This type includes all applications for processing streams of data.

## 6. Performance Evaluation and Comparison

In this section, we evaluate and analyze the performances of the four abovementioned SPCs (i.e., Cap-FIFO, Fair-Fair, Fair-FIFO, and Fair-DRF). To this end, we established a real YARN cluster by renting 31 virtual private servers (VPSs) from Linode [26], which is a virtual-private-server provider based in New Jersey. All the VPSs were located at a same datacenter in Dallas. One VPS acted as RM, and the other VPSs acted as slave nodes. Each VPS ran Ubuntu 12.04.3 LTS with 2 CPU Cores, 2 GB RAM, 48 GB SSD Storage, 3 TB Transfer, 40 Gbps Network In, and 250 Mbps Network Out [27]. Hence, total CPU capacity and total memory capacity of the YARN cluster were 60 CPU Cores and 60 GB, respectively. All the experiments were conducted on Hadoop 2.2.0 [28] with Spark 1.0.2 [29]. Table 1 lists all the parameter settings in our experiments. Other unmentioned parameters follow the default settings of YARN [28].

TABLE 1. The parameter setting of our YARN cluster

| Parameter | Value |
|---|---|
| yarn.scheduler.minimum-allocation-mb<br>(i.e., The minimum memory allocation for every container request at RM.) | 1024 MB |
| yarn.scheduler.maximum-allocation-mb<br>(i.e., The maximum memory allocation for every container request at RM.) | 2048 MB |
| yarn.scheduler.minimum-allocation-vcores<br>(i.e., The minimum virtual-CPU-core allocation for every container request.) | 1 |
| yarn.scheduler.maximum-allocation-vcores<br>(i.e., The maximum virtual-CPU-core allocation for every container request.) | 2 |
| yarn.nodemanager.resource.cpu-vcores<br>(i.e., Number of vCores that can be allocated by a node for containers.) | 2 |
| yarn.nodemanager.resource.memory-mb<br>(i.e., Amount of memory that can be allocated by a node for containers.) | 2048 MB |

Without losing the generality, we created a mixed workload to evaluate the four SPCs. Table 2 summaries the details of the workload. Note that the number of each type of applications (except for the streaming type) and the submission order of all the applications in the workload were randomly determined. The total number of applications is 94, which includes 37 two-stage applications, 28 DAG applications, 28 DCG applications, and one streaming application. The benchmarks of the two-stage applications were from [30], and the benchmarks of the other types of applications can be found in [31]. Although there is only one streaming application in the workload, its continuous running property consumes a certain amount of resources, which will be shown later. The streaming application was the first job in the workload, and it processed streams that were generated approximately every five seconds. The arrivals of the rest applications followed a Poisson distribution with the average interval 32.11 seconds and standard deviation 27.63 seconds. Regardless of application types, each of them requires one container to run their AM. Each two-stage application needs $\frac{\text{data size}}{128\text{MB}}$ containers to execute its tasks, but each of the other application types only needs two containers to run their tasks because each of them was divided into two tasks. Table 3 lists the container resource requirement for each application type of the mixed workload.

As mentioned in the Introduction, we consider the following three scenarios to evaluate each SPC. The purpose is to determine the most appropriate SPC for each queue structure and find out which queue structure is the most suitable one for mixed applications.

1. One-queue scenario: In this scenario, our YARN cluster has only one leaf queue, implying that all applications in the mixed workload will be inserted into this queue and wait for execution. It also means that this queue can use the whole resource of the cluster.

2. Separate-queue scenario: In this scenario, our YARN cluster has four leaf queues. Each queue is for a different type of applications. Hence, applications belonging to a same type will be put into a same leaf queue. The minimum and maximum capacities of each queue are 25% and 30% of the cluster resources, respectively.

3. Merged-queue scenario: Two leaf queues are in this scenario. One queue is for streaming applications with the minimum capacity of 20% and the maximum capacity of 30%. The other queue is used to put the other types of applications. Its minimum capacity and maximum capacity are 80% and 90%, respectively.

TABLE 2. The details of the mixed workload (The total number of applications is 94 with the average arrival interval 32.11 seconds and standard deviation 27.63 seconds.)

| Application type | Number | Benchmark description | Note |
|---|---|---|---|
| Two-stage | 37 | 5 wordcount applications<br>3 sort applications<br>8 grep applications<br>6 wordmean applications<br>15 wordstandarddeviation applications | Data size:<br>1 GB: 64.86%;<br>5 GB: 29.73%;<br>10 GB: 5.40% |
| DAG | 28 | 9 JavaHdfsLR applications | 192.5 KB of input size |
|  |  | 9 JavaKMeans applications | 17.31 MB of input size |
|  |  | 10 JavaPageRank applications | 14.83 MB of input size |
| DCG | 28 | 28 LiveJournalPageRank applications | 32 Bytes of input size |
| Streaming | 1 | 1 JavaQueueStream application | Data stream interval: 5 sec. |

TABLE 3. The container resource requirement for each application type of the mixed workload

| Application type | Container Resource Requirement for AM | Container Resource Requirement for each task |
|---|---|---|
| Two-stage | vCore: 1, Memory: 2048 MB | vCore: 1, Memory: 1024 MB |
| DAG | vCore: 1, Memory: 1024 MB | vCore: 1, Memory: 2048 MB |
| DCG |  |  |
| Streaming |  |  |

In addition, to comprehensively evaluate and compare the four SPCs, the following six metrics are used:

1. Workload completion rate: It shows the percentages of the workload that can be successfully completed. Note that in this paper, if an application can be successfully finished, this application is considered as complete. Otherwise, it is considered as failed. In addition, if the streaming application can continue processing streams during the entire workload execution, it is also considered as complete.

2. Cumulative workload completion rate: This metric shows the cumulative workload completion rate during the workload execution.

3. Workload turnaround time: It is the time period starting when the first application of the workload is submitted to our YARN cluster and ending when the execution of the entire workload ends (except for the streaming application), no matter if some of them are failed or not.
4. Average system load: It shows the average number of containers launched by our YARN cluster during the workload execution.
5. Streaming throughput: It is the amount of data stream that the YARN cluster can process per minute.
6. Total delay: It is the time required by the YARN cluster to schedule and process a stream of data.

To achieve a fair performance comparison, each of the four SPCs was carefully tested and evaluated for five times, no matter which scenario was employed.

**6.1 The one-queue scenario**

In this subsection, we show the execution performances of the four SPCs in the one-queue scenario. Fig. 3 shows that when the four SPCs were individually employed to run the workload, some applications of the workload could not finish due to failing to get containers. None of them can achieve 100% of completion rate. The key reason is that the container-based resource allocation utilized by YARN causes that no slave at the same moment has sufficient available resources to launch a desired container for an application. We call this a resource fragmentation problem. In our experiment, some containers request 1024 MB of memory, and some other request 2048 MB of memory. Hence, if an application needs a container with 2048 MB of memory but no slave can afford it at the moment, this application cannot be executed.

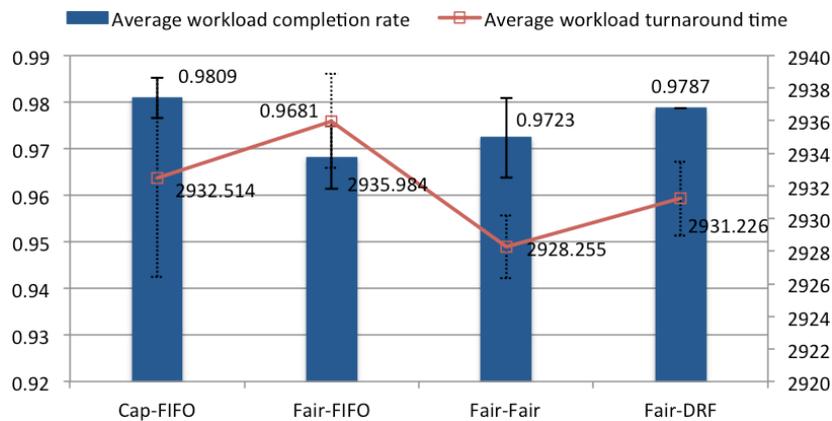

Fig. 3. The average workload completion rates and average workload turnaround time of the four SPCs in the one-queue scenario.

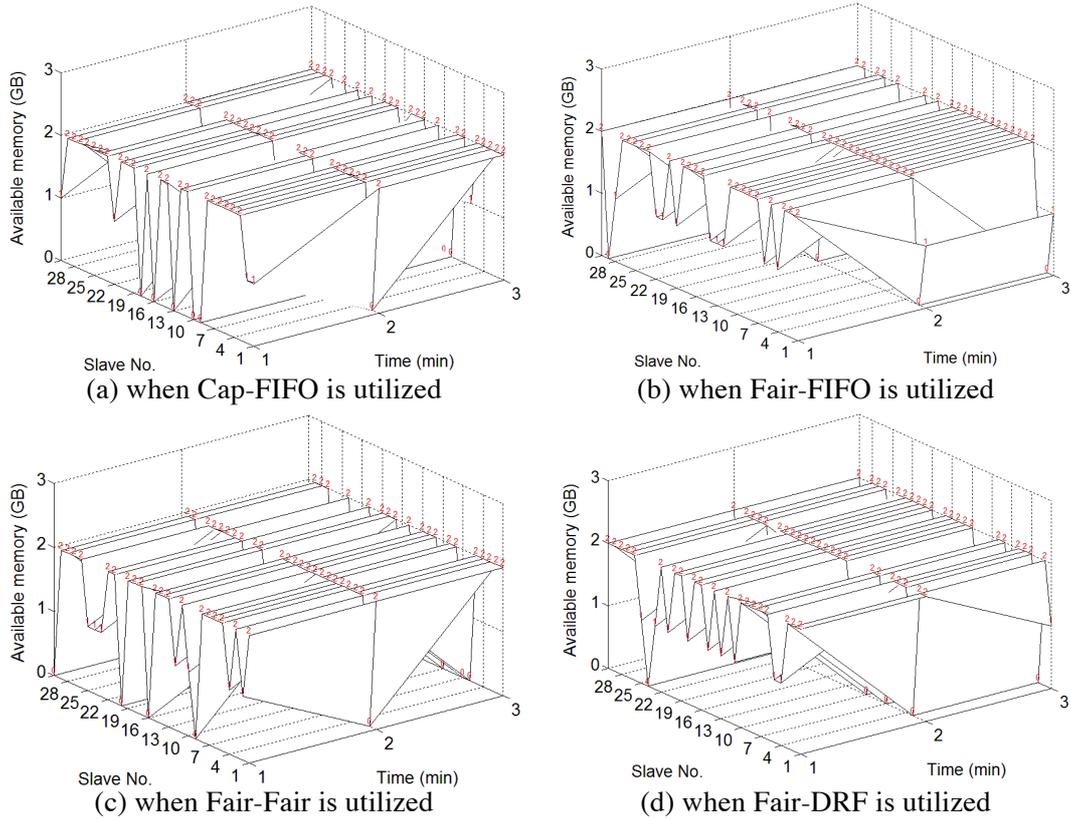

Fig. 4. The available memory of all slaves at the first three minutes of the workload execution.

Although all SPCs suffer from the resource fragmentation problem, Cap-FIFO provided the highest workload completion rate (about 98.09%). This is because Cap-FIFO tends to launch a new container from a used slave as long as the remaining resource of the slave is sufficient to create the container. This property can be seen by comparing Fig. 4a with Figs. 4b, 4c, and 4d. Fig. 4a shows that when Cap-FIFO was tested, five slaves had no memory available and four slaves had 1 GB of memory available at the first minute of the workload execution. However, Figs. 4b, 4c, and 4d illustrate that when the other three SPCs were tested, more than four slaves had 1 GB of memory available at the first minute, meaning that these slaves cannot create a container for any applications that need a container with 2 GB of memory. Based on the above results, we can see that the container launch manner used by Cap-FIFO is more gentle, which mitigates the resource fragmentation problem and therefore improves workload completion rate. Due to the same reason and the resource fragmentation problem, both Fair-FIFO and Fair-Fair had lower completion rates than Cap-FIFO. But we found that the workload completion rate of Fair-DRF was not significantly impacted, implying that the DRF policy used by Fair-DRF can also mitigate the abovementioned problems.

Fig. 3 also depicts the average workload turnaround time of the four SPCs. Although Fair-Fair led to the shortest average workload turnaround time, it is not a good SPC for the one-queue scenario since its completion rate was lower than those of Cap-FIFO and Fair-DRF. Based on the average workload completion rate, average workload turnaround time, and standard deviation shown in Fig. 3, we can see that Fair-DRF performs the best, whereas Fair-FIFO performs the worst.

Fig. 5 illustrates the average cumulative workload completion rates of the four SPCs during the workload execution. We can see that the four curves are almost overlapped, implying that all SPCs have similar workload execution speeds.

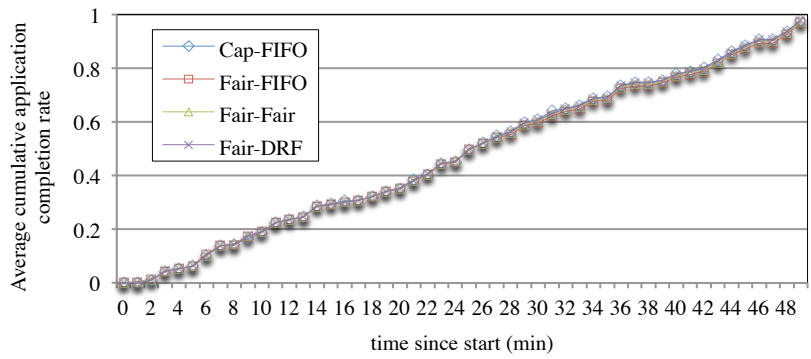

Fig. 5. The average cumulative workload completion rates of the four SPCs in the one-queue scenario.

Fig. 6 illustrates the average system load of the four SPCs. When Cap-FIFO was tested, the cluster in average launched 1112 containers to perform the workload. This value is lower than those of the other three SPCs, implying that Cap-FIFO saves more containers than the other SPCs.

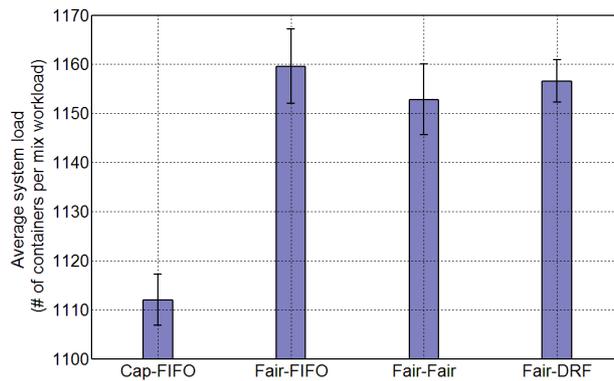

Fig. 6. The average system load when the four SPCs were individually employed in the one-queue scenario.

Fig. 7 shows the average streaming throughput of the four SPCs. At the beginning of the workload execution, all SPCs could process more than 12 streams of data per minute. However, as more applications of the workload were submitted to the cluster, all SPCs' streaming throughputs reduced. Nevertheless, we still can see that Fair-DRF provides a slightly higher throughput than the others. Fig. 8 illustrates the average total delays of the four SPCs. No matter which SPC was utilized, the differences among their average total delays at the first quartile, median, and the third quartile were insignificant, and their standard deviations were similar to each other, implying that these four SPCs have indistinguishable performance in terms of total delay.

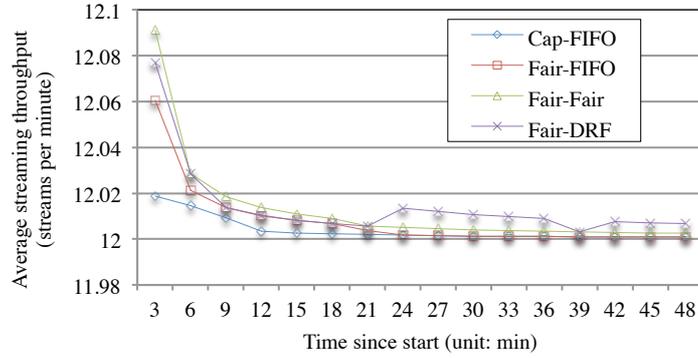
Fig. 7. The average streaming throughputs of the four SPCs in the one-queue scenario.

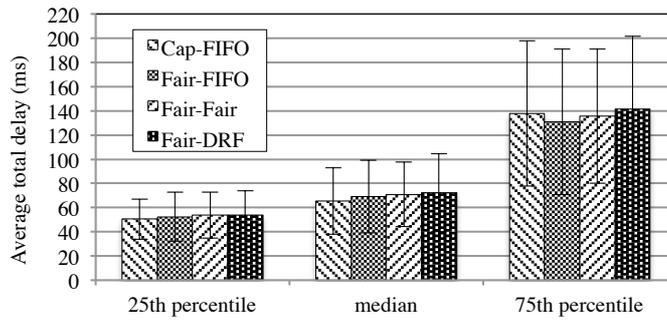
Fig. 8. The average total delays of the four SPCs in the one-queue scenario.

Based on the results shown from Figs. 3 to 8, we draw the following conclusions: If applications execution performance is the most important concern, Fair-DRF is the most recommended SPC for the one-queue scenario because of its good performance in terms of workload completion rate, workload turnaround time, and streaming throughput. However, if we only consider resource-usage efficiency, Cap-FIFO is suggested since it uses less containers than the other SPCs.

**6.2 The separate-queue scenario**

In this subsection, we evaluate how the four SPCs perform in the separate-queue scenario. Note that all SPCs had close cumulative workload completion rates during the workload execution, so the results were not depicted to save paper space.

Fig. 9 shows that Cap-FIFO provided the highest workload completion rate and the smallest standard deviation among all SPCs. But Cap-FIFO in this scenario could not complete as many applications as it could in the one-queue scenario (please compare Fig. 9 with Fig. 3). The main reasons are two. First, each queue in the separate-queue scenario can use at most 30% of the cluster resources. Second, the streaming application always occupies 5 vCores and 5120 MB, i.e., 8.3% of the cluster resources. Hence, the other three queues for the two-stage, DAG, and DCG applications can only utilize at most 30% of the cluster resources individually, no matter that they need more. Compared with the one-queue scenario, the resources available for the two-stage, DAG, and DCG applications in the separate-queue scenario reduced, and hence caused more application faults.

The above phenomenon not only occurs when Cap-FIFO was utilized, but also happens when the other three SPCs were tested. By comparing Fig. 9 and Fig. 3, we can see that the workload completion rates of Fair-FIFO, Fair-Fair, and Fair-DRF in the separate-queue scenario were not as high as those in the one-queue scenario. The situation is even worse for Fair-DRF since its average completion rate dropped to 95.96% with a very large standard deviation, implying that Fair-DRF is inappropriate for the separate-queue scenario.

Fig. 9 also reveals that the average workload turnaround time of Cap-FIFO was longer than those of the three SPCs, and its corresponding standard deviation was the largest despite its high workload completion rate. On the other hand, even though Fair-FIFO's workload completion rate was the second best (see Fig. 9), its workload turnaround time was shorter than Cap-FIFO's. Hence, from the perspective of both workload completion rate and workload turnaround time, Fair-FIFO is more suitable for the separate-queue scenario. By comparing Fig. 9 with Fig. 3, we see that the four SPCs in the separate-queue scenario led to a slightly longer workload turnaround time than they did in the one-queue scenario, implying that employing four leaf queues is no better than employing one leaf queue.

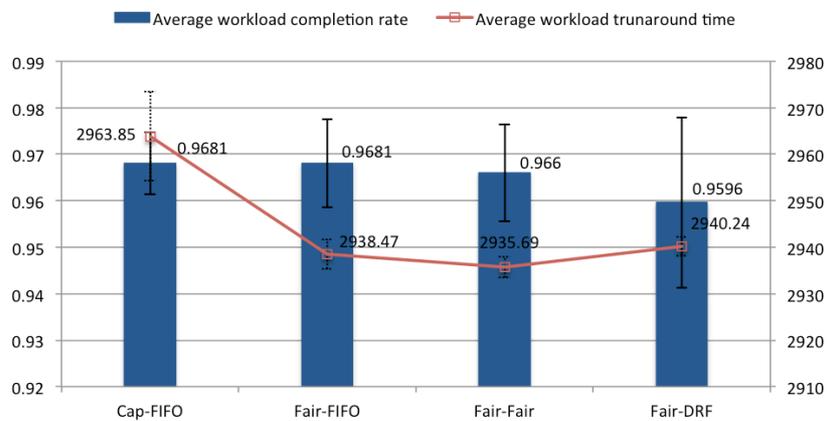
Fig. 9. The average workload completion rates and average workload turnaround time of the four SPCs in the separate-queue scenario.

Fig. 10 shows the average system load caused by the four SPCs. By comparing Fig. 10 with Fig. 6, we find that all SPCs in the separate-queue scenario led to a lower average system load than they did in the one-queue scenario. This is because the resource fragmentation problem and the capacity limit for each queue disallow these SPCs to launch more containers to run the workload, which therefore impacts workload completion rate.

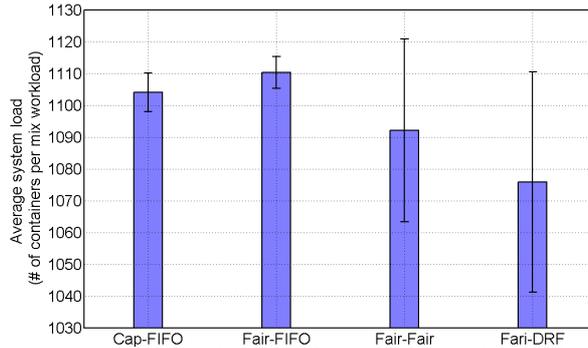

Fig. 10. The average system load when the four SPCs are individually employed in the separate-queue scenario.

Fig. 11 illustrates the average streaming throughput of the four SPCs in the separate-queue scenario. We can see that the streaming throughputs of Fair-Fair and Fair-DRF were both less than 12 streams per minute. The key reason is that when these two SPCs were employed in the separate-queue scenario, the resources allocated to the streaming queue were mostly occupied by the other applications. More clearly, each queue for the two-stage, DAG, and DCG applications used 30% of the cluster resources, and the streaming queue only used 10%. Due to such resource competition in Fair-Fair and Fair-DRF, the streaming application was unable to provide a good throughput.

Fig. 12 shows the average total delays of the four SPCs. Since Fair-Fair and Fair-DRF had low streaming throughput, we can see that their average total delays at the median and third quartile were slightly longer than those of Cap-FIFO and Fair-FIFO.

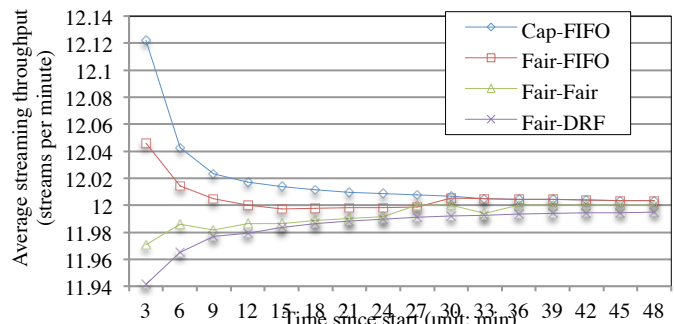

Fig. 11. The average streaming throughputs of the four SPCs in the separate-queue scenario.

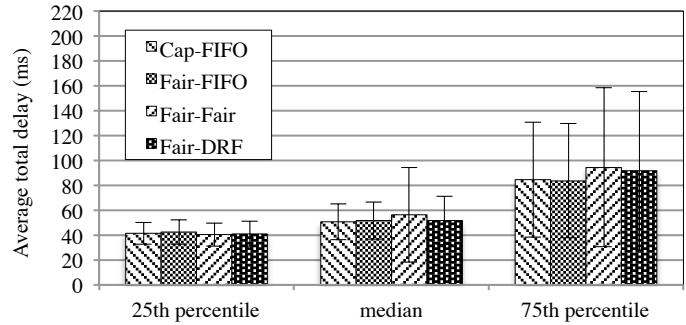

Fig. 12. The average total delays of the four SPCs in the separate-queue scenario.

Based on the results shown in Figs. 9 to 12, we conclude that Fair-FIFO is the best in the separate-queue scenario from the perspective of applications execution performance. In addition, if we compare the performances of the four SPCs in the separate-queue scenario with those in the one-queue scenario, we found that all SPCs have better performance in the one-queue scenario since they did not suffer from the resource shortage problem caused by the capacity limitation of each leaf queue in the separate-queue scenario. Hence, using one queue to organize the mixed applications is better than using four queues.

**6.3 The merged-queue scenario**

In order to study whether these SPCs can perform better than they do in the previous scenarios, in this subsection, we evaluate them in the merged-queue scenario. Fig. 13 illustrates that both Cap-FIFO and Fair-DRF achieved the highest completion rate (about 97.87%), and both Fair-FIFO and Fair-Fair provided the second best completion rate (about 97.52%). By comparing Fig. 13 with Fig. 3, we can see that the workload completion rates of all SPCs (except for Cap-FIFO) increased in the merged-queue scenario, implying that for these SPCs separating the streaming application and the other three types of applications into two different queues enables more applications of the workload to be successfully completed. The key reason is that the resource used by the streaming application was at most 8.3% of the entire cluster resources. Hence, the rest resources allocated to the streaming queue could be freely competed by the other types of applications.

Although Cap-FIFO performed as good as Fair-DRF in terms of workload completion rate, its workload turnaround time was slightly longer than that of Fair-DRF (please see Fig. 13). Similarly, even though Fair-FIFO had the same completion rate as Fair-Fair, its workload turnaround time was slightly longer than that of Fair-Fair. By comparing Fig. 13 with Fig. 3, it is clear that all SPCs led to a slightly shorter workload turnaround time in the merged-queue scenario. Hence, we can conclude that the merged-queue scenario not only improves the workload completion rates for almost all SPCs, but also shortens their workload turnaround time.

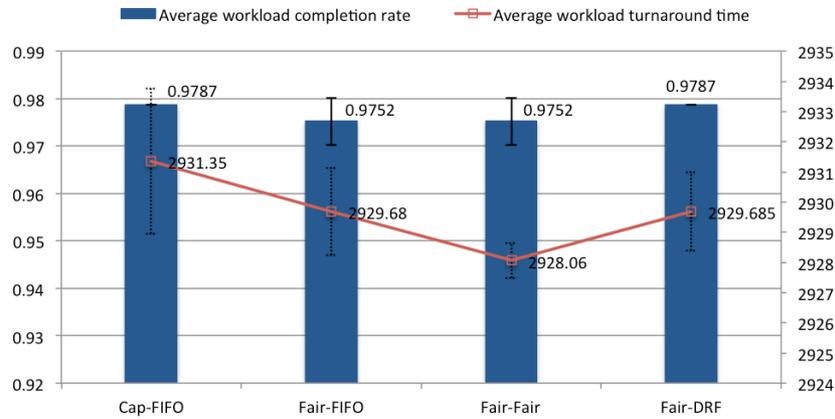
Fig. 13. The average workload completion rates and average workload turnaround time of the four SPCs in the merged-queue scenario.

Fig. 14 illustrates that Cap-FIFO has the lowest average system load among all SPCs, and it was not affected by all the three scenarios, implying that Cap-FIFO is the most efficient in terms of container usage. However, the same situation does not occur when the other SPCs were tested. We can see that the average system loads of Fair-Fair and Fiar-DRF slightly increased in the merged-queue scenario (please compare Fig. 14 with Fig. 6). This is because in the merged-queue scenario these SPCs could complete more applications of the workload, and hence the total number of containers used to run the workload increased.

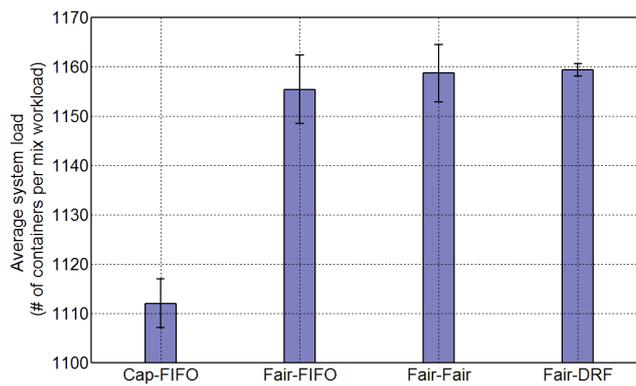
Fig. 14. The average system load when the four SPCs are individually employed in the merged-queue scenario.

Fig. 15 illustrates the streaming throughput of all SPCs. We can see that only Fair-DRF has the streaming throughput less than 12 streams per minute since the resource allocated to the streaming queue under this SPC were mostly used by other applications. However, the same problem was mitigated when Fair-Fair was utilized. By comparing Fig. 15 with Fig. 11, it is clear that the streaming throughput of Fair-Fair improved when the merged-queue scenario was employed. Fig. 16 illustrates the average total delays of the four SPCs. Since Fair-DRF's streaming throughput was not good during most time of the workload execution, its average total delays and standard deviation were slightly higher than those of the other SPCs.

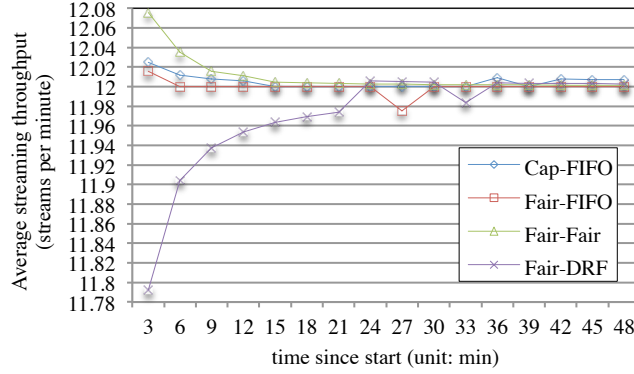

Fig. 15. The average streaming throughputs of the four SPCs in the merged-queue scenario.

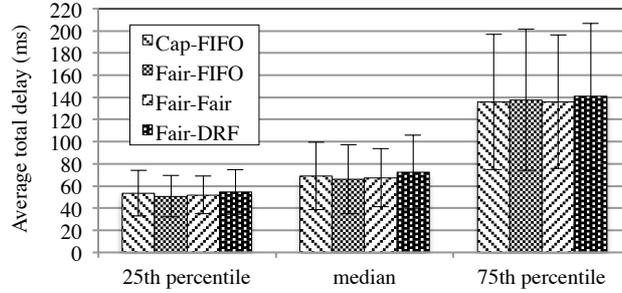

Fig. 16. The average total delays of the four SPCs in the merged-queue scenario.

Based on the experiment results shown in Figs. 13 to 16, we conclude that Cap-FIFO and Fair-DRF are both recommended for the merged-queue scenario due to they can achieve high workload completion rate. However, if workload turnaround time is further considered, Fair-DRF performs slightly better than Cap-FIFO. On the other hand, if streaming throughput is considered, Cap-FIFO is slightly better than Fair-DRF. But from the viewpoint of resource-usage efficiency, Cap-FIFO is still the best.

TABLE 4. The most recommended SPC when different metrics and different queue-structure scenarios are considered

| Metric / Scenario | Application execution performance | Resource-usage efficiency |
|---|---|---|
| One-queue scenario | Fair-DRF | Cap-FIFO |
| Separate-queue scenario | Fair-FIFO | Fair-DRF |
| Merged-queue scenario | Cap-FIFO and Fair-DRF | Cap-FIFO |

## 7. Conclusions and Future Work

In this paper, we have surveyed the four SPCs and four application types supported by YARN. To fully understand the performance impacts of the four SPCs on mixed application types, we conducted extensive experiments by considering not only a workload comprising mixed application types, but also three different queue-structure scenarios (i.e., one-queue scenario, separate-queue scenario, and merged-queue scenario). Based on the experimental results, we draw the following conclusions and summarize our suggestions in Table 4.

1. Fair-DRF is the best choice for the one-queue scenario since it leads to a higher workload completion rate and shorter workload turnaround time as compared with the other three SPCs.
2. Fair-FIFO is the most recommended SPC for the separate-queue scenario due to its good performance in terms of both workload completion rate and workload turnaround time.
3. Cap-FIFO and Fair-DRF are both appropriate for the merged-queue scenario. However, Cap-FIFO is slightly better than Fair-DRF in streaming throughput and resource-usage efficiency, whereas Fair-DRF is slightly better than Cap-FIFO in workload turnaround time.
4. From the viewpoint of resource-usage efficiency, Cap-FIFO performs the best in the one-queue and merged-queue scenarios since the total number of containers launched by Cap-FIFO to execute the workload is less that those launched by the other three SPCs.

If we take the experimental results of all the scenarios into consideration, it is apparent that employing the merged-queue scenario is the best choice for all SPCs since it enables almost all SPCs to achieve high workload completion rate and shorten workload turnaround time. On the contrary, utilizing the separate-queue scenario is not recommended since it worsens workload completion rates and prolongs workload turnaround time for almost all SPCs.

Our future work is study how various combinations of applications of a workload impact the above SPCs and further to propose a new scheduler for YARN such that the resource fragmentation problem can be mitigated and workload completion rate can be improved.

**Acknowledgement**

The work was supported by the scholarship of the Sandwich Programme supported by Ministry of Science and Technology, Taiwan and Deutscher Akademischer Austausch Dienst (DAAD) under Grants NSC 102-2911-I-100-524 and NSC 101-2911-I-009-020-2. The authors also want to thank the anonymous reviewers for their reviews and suggestions to this paper.